\documentstyle[aps,preprint,12pt]{revtex}

\draft
\title{\flushright{\small AEI--2002--071}\\ \centerline{Constraints Algebra and Equations of Motion in}\centerline{Bohmian Interpretation of Quantum Gravity}}
\author{ALI SHOJAI\thanks{Email:  SHOJAI@IPM.IR}}
\address{Physics  Department, Tehran University, Tehran, Iran} \address{and}  \address{Institute for
Studies in Theoretical  Physics and Mathematics,  P.O.Box  19395-5531,  Tehran, Iran}  \address{and}
\address{MPI f.  Gravitationsphysik,  Albert--Einstein--Institut, Am M\"uhlenberg 1, 14476 Golm near
Potsdam, Germany}
\author{FATIMAH SHOJAI\thanks{Email:  FATIMAH@IPM.IR}}
\address{Institute  for Studies in Theoretical Physics and Mathematics,  P.O.Box 19395-5531, Tehran,
Iran} \address{and} \address{MPI f.  Gravitationsphysik, Albert--Einstein--Institut, Am M\"uhlenberg
1, 14476 Golm near Potsdam, Germany}
\begin{document}
\maketitle
\begin{abstract} 
It is shown that introducing the quantum effects using deBroglie--Bohm theory in the canonical formulation of gravity would change the constraints algebra. The new algebra is derived and shown that it is the clear projection of general coordinate transformations to the spatial and temporal diffeomorphisms. The quantum Einstein's equations are derived and it is shown that they are manifestly covariant under the above diffeomorphisms, as it would be.
\end{abstract}
\section{Introduction}
The issue of constraints algebra is an important problem in canonical quantum gravity\footnote{In this paper we are dealing with WDW approach to canonical quantum gravity and Ashtekar's new variables are not considered.}. Unfortunately the constraints algebra is not a true Lie algebra because, dynamical variables appear in the structure constants. More important is that the poisson bracket of Hamiltonian constraint with itself is not zero. Therefore the algebra is not a clear projection of 4--diffeomorphism in foliated space--time\cite{kuc,mar1}. Using a fixed embedding restricts the symmetry of the theory to a subgroup of general coordinate transformations, invariance under spatial and temporal reparametrizations. Spatial reparametrization invariance is explicitly present in the algebra, but temporal reparametrization invariance is not present explicitly.
Recently\cite{kuc,mar1,mar2} some modifications are done to have the appropriate algebra.

On the other hand, Bohm's causal interpretation of quantum mechanics (see e.g. \cite{hor}) may provide a better understanding of some aspects of quantum gravity in comparison to the Copenhagen interpretation\cite{gold}. We shall discuss some aspects of Bohmian interpretation in the next section. 

In Bohm's theory any system has a definite trajectory and any quantity is a c-number not an operator. One must modify the classical hamiltonian by adding the quantum potential. In this way the constraints algebra could be evaluated and may differ from the classical one. This is what we shall do in section III. 

In addition since in Bohm's interpretation of quantum mechanics the system has a well--defined trajectory, the evolution equations of the metric (quantum Einstein's equations) can be obtained. This is done in section IV.
\section{Why Bohmian Quantum Gravity?}
In this work we use the Bohmian interpretation of canonical quantum gravity, because it has some useful aspects\cite{hor,hor1}. Some of them are:
\begin{itemize}
\item It leads to time evolution of the dynamical variables whether the wave function depends on time or not.
Therefore we have not the time problem in Bohmian quantum gravity.
\item 	Bohm's theory describes a single system, unlike the Copenhagen interpretation of quantum theory, which does not tell anything about a single system. About an ensemble of the system both interpretations are equivalent. This is because of the specific form of Bohm's equations of motion. They are the Bohm-Hamilton-Jacobi equation and the conservation equation of probability. These equations can be transformed to the Schr\"odinger equation by some canonical transformation.

This aspect is useful in quantum cosmology where the system is the universe, and an ensemble of it does not exist.
Therefore we have not here the conceptual problem of the meaning of the universe's wave function in quantum cosmology.
\item Normalization of the wave function is needed only for the probabilistic description. Here there is no need to normalize the wave function for a single system.
\item The classical limit has a well-defined meaning. When the quantum potential is less than the classical potential and the quantum force is less than the classical force we are in the classical domain.
\item There is no need to separate the classical observer and the quantum system in the measurement problem. In the Bohmian picture of the measurement process we have two interacting systems, the system and the observer. After the interaction takes place, the wave function of the system is reduced in a causal way.
\end{itemize}
It must be noted that the same statistical results for the Copenhagen and Bohmian interpretations doesn't mean that the two theories are equivalent. They are different in physical concepts. The most important difference is that in Bohmian interpretation one deals with trajectories. This can lead to new concepts. E.g., one can evaluate the tunneling time of the particle through a potential barrier in the non--relativistic quantum mechanics. This is a concept that has not a clear meaning in the Copenhagen interpretation\cite{hor,cush}. 

Till now Bohmian interpretation of WDW quantum gravity and cosmology have given some physical results that could be found in the literature:
\begin{itemize}
\item In Bohmian quantum cosmology the quantum force can remove the big bang singularity, because it can behave as a repulsive force. This has been shown for a radiation dominated universe in \cite{bar,ss1}, for a dust filled flat FRW universe in \cite{bar} and for FRW model with a minimal massless scalar field in \cite{kim}.
\item The quantum force may be present in large scales because the quantum effects of quantum potential are independent of the scale\cite{col}.
This behaviour can be seen for flat FRW universe with dust and a conformal invariant scalar field in \cite{lem}.
\item One can find the graceful exit behaviour in super inflation model in a super string cosmology. The evolution begins with inflation and smoothly changes to the decelerating expansion, without any singularity in the transition\cite{net}.
For a more detailed discussion of deBroglie--Bohm interpretation of quantum super string cosmology, pre--big bang inflation and graceful exit problem considering various classes of wave packet see \cite{mart}.
\item Real time tunneling can be occurred in the classically forbidden regions, through quantum potential. For this effect in a closed deSitter universe in 2+1 dimensions see \cite{ken}.
\item In a generalized geometric picture of Bohm's interpretation, one can unify the quantum effects and gravity\cite{ss}.
\end{itemize} 
Some useful questions that can be answered in Bohm's interpretation are that, what is the Constraints algebra?, is it a correct Lie algebra?, is it possible to see explicitly the symmetries at the level of equation of motion of metric? 
We shall discuss these topics in details in the next sections.

\section{Quantum constraints algebra}
Using the ADM decomposition and assuming that there is no matter field, the lagrangian density of general relativity is ($16\pi G=1$):
\begin{equation}
{\cal L}=\sqrt{-g}{\cal R}=\sqrt{q}N\left ( ^{(3)}{\cal R}+\text{trace} (K^2) -(\text{trace} K)^2\right )
\end{equation}
where $^{(3)}{\cal R}$ is the 3-dimensional Ricci scalar, $K_{ij}$ is the extrinsic curvature and $q_{ij}$ is the induced spatial metric.
The canonical momentum of the 3-metric is given by:
\begin{equation}
p^{ij}=\frac{\partial {\cal L}}{\partial \dot{q}_{ij}}=\sqrt{q}\left ( K^{ij}-q^{ij}\text{trace} K\right )
\end{equation}
The classical Hamiltonian is:
\begin{equation}
H=\int d^3 x {\cal H}
\end{equation}
where the hamiltonian density is:
\begin{equation}
{\cal H}=\sqrt{q}(NC+N^iC_i)
\end{equation}
where $N$ and $N_i$ are lapse and shift functions.
And the constraints are given by:
\begin{equation}
C=-^{(3)}{\cal R}+\frac{1}{q}\left ( \text{trace} (p^2)-\frac{1}{2}(\text{trace} p)^2\right )=-2G_{\mu\nu}n^\mu n^\nu
\label{a}
\end{equation}
and
\begin{equation}
C_i=-2^{(3)}\nabla^j\left (\frac{p_{ij}}{\sqrt{q}}\right )=-2G_{\mu i}n^\mu
\label{b}
\end{equation}
in which $n^\mu$ is the normal vector to the spatial hyper surfaces given by $n^\mu=(1/N,-\vec{N}/N)$.
It is well known that in Bohm's theory one must add the quantum potential to Hamiltonian to get the correct equations of motion\cite{hor}, that is:  
\begin{equation}
H\longrightarrow H+Q
\end{equation}
or
\begin{equation}
{\cal H}\longrightarrow {\cal H}+{\cal Q}
\end{equation}
where
\begin{equation}
Q=\int d^3x {\cal Q}
\end{equation}
The quantum potential is given by:
\begin{equation}
{\cal Q}=\hbar^2NqG_{ijkl}\frac{1}{|\Psi|}\frac{\delta^2|\Psi|}{\delta q_{ij}\delta q_{kl}}
\end{equation}
where $G_{ijkl}$ is the super space metric and  $\Psi$ is the wave function, satisfying the WDW equation.
This means that one must modify the classical constraints in this way:  
\begin{equation}
C\longrightarrow C+\frac{{\cal Q}}{\sqrt{q}N}
\end{equation}
and
\begin{equation}
C_i\longrightarrow C_i
\end{equation}
Now we can evaluate the constraint algebra. We use the integrated forms of the constraints defined as:
\begin{equation}
C(N)=\int d^3x \sqrt{q}NC
\end{equation}
and
\begin{equation}
\widetilde{C}(\vec{N})=\int d^3x \sqrt{q}N^iC_i
\end{equation}
We shall get\cite{pra}:
\begin{equation}
\left \{ \widetilde{C}(\vec{N}), \widetilde{C}(\vec{N}')\right \}=\widetilde{C}(\vec{N}\cdot\nabla
\vec{N}'-\vec{N}'\cdot\nabla\vec{N})
\label{c1}
\end{equation}
\begin{equation}
\left \{ \widetilde{C}(\vec{N}), C(N)\right \}=C(\vec{N}\cdot\nabla N)
\label{c2}
\end{equation}
\begin{equation}
\left \{ C(N), C(N')\right \}\sim 0
\label{lanati}
\end{equation}
The first one, the 3-diffeomorphism sub algebra, has no change with respect to the classical one. 
And the second, representing the fact that the Hamiltonian constraint is a scalar under 3-diffeomorphism, is also the same as in the classical case. 
In the third one, the quantum potential changes the Hamiltonian constraint algebra dramatically. 
This Poisson bracket is weakly zero.
To see this, let us to write the poisson bracket explicitly:
\[ 
\left \{ C(N), C(N')\right \}=\int d^3z \sqrt{q(z)} \left ( \frac{\delta C(N)}{\delta q_{ij}(z)} \frac{\delta C(N')}{\delta p^{ij}(z)}- \frac{\delta C(N)}{\delta p^{ij}(z)} \frac{\delta C(N')}{\delta q_{ij}(z)}\right )=
\]
\begin{equation}
\widetilde{C}(N\vec{\nabla}N'-N'\vec{\nabla}N)+2\int d^3z d^3x \sqrt{q(z)}G_{ijkl}(z) p^{kl}(z) \left ( -N(z)N'(x)+N(x)N'(z)\right ) \frac{\delta ({\cal Q}/(\sqrt{q}N))}{\delta q_{ij}(z)}
\label{xax}
\end{equation}
To simplify this, one only needs to differentiate the Bohm--Hamilton--Jacobi equation\cite{pra} that gives us:
\begin{equation}
\frac{1}{N}\frac{\delta}{\delta q_{ij}}\frac{{\cal Q}}{\sqrt{q}}=
\frac{3}{4\sqrt{q}}q_{kl}p^{ij}p^{kl}\delta(x-z)-\frac{\sqrt{q}}{2}q^{ij}
\left ( ^{(3)}{\cal R}-2\Lambda \right ) \delta(x-z)
-\sqrt{q}\frac{\delta ^{(3)}{\cal R}}{\delta q_{ij}}
\label{yay}
\end{equation}
and use it in the evaluation of the poisson bracket. The result is the relation (\ref{lanati}). The complete derivation is introduced in the appendix.
We see that the existence of quantum potential tells us the quantum algebra is 3-diffeomorphism times an abelian sub algebra but the only difference with the ref. \cite{mar1}is that this algebra is weakly closed. 
Although these are the results that obtained in our previous work \cite{pra}, here we present the explicit physical meaning of the constraint algebra. That is, here we see that the algebra (\ref{c1})--(\ref{lanati}) is a clear projection of general coordinate transformations to the spatial and temporal diffeomorphisms. We shall see in the next section that the equations of motion are invariant under such transformations. Also in \cite{net1} for a special form of quantum potential the same result is obtained. 

The important point here is that although the form of quantum potential depends on the regularization and ordering, but in the quantum constraints algebra the form of quantum potential is not important. They are in fact correct independent of the definition of quantum potential.

It must be noted here that although we have derived these poisson brackets for pure gravity, but the inclusion of matter doesn't change anything and we have again the same algebra.
\section{QUANTUM EINSTEIN EQUATIONS}
Now we can derive the quantum corrections to Einstein's equations. For the dynamical part consider the Hamilton equations:
\begin{equation}
\dot{q}^{ij}=\left \{ H, q^{ij}\right \}
\end{equation}
and
\begin{equation}
\dot{p}_{ij}=\left \{ H, p_{ij}\right \}
\end{equation}
which yield to the following quantum equations:
\begin{equation}
\dot{q}_{ij}=\frac{2}{\sqrt{q}}N \left ( p_{ij}-\frac{1}{2}p_k^kq_{ij}\right ) +2 ^{(3)}\nabla_{[i}N_{j]}
\end{equation}
and
\[
\dot{p}^{ij}=-N\sqrt{q}\left ( ^{(3)}{\cal R}^{ij}-\frac{1}{2}\ ^{(3)}{\cal R}q^{ij}\right )
+\frac{N}{2\sqrt{q}}q^{ij}\left ( p^{ab}p_{ab}-\frac{1}{2}(p^a_a)^2\right )
\]
\[
-\frac{2N}{\sqrt{q}}\left ( p^{ia}p^j_a-\frac{1}{2}p^a_a p^{ij}\right )+ \sqrt{q}
\left ( \nabla^i \nabla^j N -q^{ij}\ ^{(3)}\nabla^a\ ^{(3)}\nabla_a N\right )
\]
\begin{equation}
+\sqrt{q} ^{(3)}\nabla_a \left ( \frac{N^a}{\sqrt{q}}p^{ij}\right ) -2p^{a[i}\ ^{(3)}\nabla_a N^{j]}
-\sqrt{q}\frac{\delta Q}{\delta q_{ij}}
\end{equation}
Combining these two equations, we get after some calculations:
\begin{equation}
{\cal G}^{ij}=-\frac{1}{N}\frac{\delta Q}{\delta q_{ij}}
\label{vayyyy}
\end{equation}
which means that the quantum force modifies the dynamical parts of Einstein's equations.
For the non dynamical parts we use the constraint relations (\ref{a}), (\ref{b}) obtaining:
\begin{equation}
{\cal G}^{00}=\frac{{\cal Q}}{2N^3\sqrt{q}}
\end{equation}
and
\begin{equation}
{\cal G}^{0i}=-\frac{{\cal Q}}{2N^3\sqrt{q}}N^i
\end{equation}
These last two equations can be written as:
\begin{equation}
{\cal G}^{0\mu}=\frac{{\cal Q}}{2\sqrt{-g}}g^{0\mu}
\end{equation}
Hence the non dynamical parts are modified by quantum potential.

A problem that must be addressed here is that for a reparametrization invariant theory there is the possibility that the equations obtained by the Hamiltonian may differ from those given by the phase of the wave function and the guidance formula. We can see that there is no difference here. Let us to write the Bohmian Hamilton--Jacobi equation (see e.g. \cite{hor1}) by decomposing the phase part of the WDW equation. We have:
\begin{equation}
G_{ijkl}\frac{\delta S}{\delta q_{ij}}\frac{\delta S}{\delta q_{kl}}-\sqrt{q}\left (      	  ^{(3)}{\cal R} -{\cal Q}\right )=0
\end{equation}
where $S$ is the phase of the WDW wave function. In order to get the equation of motion one must differentiate the above Hamilton--Jacobi equation with respect to $q_{ab}$ and use the guidance formula $p^{kl}\equiv \sqrt{q}(K^{kl}-q^{kl}K)=\delta S / \delta q_{kl}$. After some tedious calculations one arrives at the same dynamical equation (\ref{vayyyy}). This shows that the evolution generated by the hamiltonian is compatible with the guidance formula, that is to say the poisson bracket of the hamiltonian and the guidance relation ($\chi^{kl}=p^{kl}-\delta S / \delta q_{kl}$) is zero. This can be evaluated explicitly and it equals to zero weakly. So everything is consistent.
 
Now we can show explicitly that these modified Einstein equations are covariant under spatial and temporal diffeomorphisms.
First consider the transformation in which $t\rightarrow t'=f(t)$ and $\vec{x}$ unchanged, we have:
\begin{equation}
q'_{ij}=q_{ij}
\end{equation}
\begin{equation}
N'_i=\frac{df}{dt} N_i
\end{equation}
\begin{equation}
N'=\frac{df}{dt}N
\end{equation}
After substituting these in modified Einstein's equations, we see that the right hand side transforms as a second rank tensor under time reparametrization.

Second consider the transformation $\vec{x}\rightarrow\vec{x}'=\vec{g}(\vec{x})$ and t unchanged, we have:
\begin{equation}
q'_{ij}=\frac{\partial x^l}{\partial x'^i}\frac{\partial x^m}{\partial x'^j}q_{lm}
\end{equation}
\begin{equation}
N'_i=\frac{\partial x^l}{\partial x'^i}N_l
\end{equation}
\begin{equation}
N'=N
\end{equation}
Again the right hand side of modified Einstein's equations is second rank tensor under spatial 3-diffeomorphism.

Inclusion of matter fields is straightforward.
Only one must add the matter quantum potential to the total quantum potential and introduce the energy momentum tensor in the equations:
\begin{equation}
{\cal G}^{ij}=-\kappa {\cal T}^{ij}-\frac{1}{N}\frac{\delta (Q_G+Q_m)}{\delta g_{ij}}
\label{c}
\end{equation}
\begin{equation}
{\cal G}^{0\mu}=-\kappa{\cal T}^{0\mu}+\frac{{\cal Q}_G+{\cal Q}_m}{2\sqrt{-g}}g^{0\mu}
\label{d}
\end{equation}
where the quantum potential of matter is defined in terms of the wave function just as the gravitational quantum potential:
\begin{equation}
{\cal Q}_m=\hbar^2\frac{N\sqrt{q}}{2}\frac{1}{|\Psi|}\frac{\delta^2|\Psi|}{\delta\phi^2}
\end{equation}
where $\phi$ is the matter field, and remember that:
\begin{equation}
{\cal Q}_G=\hbar^2NqG_{ijkl}\frac{1}{|\Psi|}\frac{\delta^2|\Psi|}{\delta q_{ij}\delta q_{kl}}
\end{equation}
and:
\begin{equation}
Q_G=\int d^3x {\cal Q}_G
\end{equation}
\begin{equation}
Q_m=\int d^3x {\cal Q}_m
\end{equation}

The equations (\ref{c}) and (\ref{d}) are the Bohm-Einstein equations which are in fact the quantum version of Einstein equations. As it is discussed previously, regularization only affects on the quantum potential, so for any regularization, the quantum Einstein's equations are the same.

As we have shown explicitly, they are invariant under temporal $\otimes$ spatial diffeomorphisms. Note that they can be written in the form:
\begin{equation}
{\cal G}^{\mu\nu}=-\kappa{\cal T}^{\mu\nu}+{\cal S}^{\mu\nu}
\label{e}
\end{equation}
where:
\begin{equation}
{\cal S}^{0\mu}=\frac{{\cal Q}_G+{\cal Q}_m}{2\sqrt{-g}}g^{0\mu}=\frac{{\cal Q}}{2\sqrt{-g}}g^{0\mu}
\end{equation}
\begin{equation}
{\cal S}^{ij}=-\frac{1}{N}\frac{\delta (Q_G+Q_m)}{\delta g_{ij}}=-\frac{1}{N}\frac{\delta Q}{\delta g_{ij}}
\end{equation}
${\cal S}^{\mu\nu}$ is the quantum correction tensor. It is a tensor under the temporal $\otimes$ spatial diffeomorphisms subgroup of the general coordinate transformations. Note that we must confine ourselves to this subgroup. This is because we must solve the WDW equation (which is written in ADM decomposition) to get the wave function and then the quantum potential. If one wishes to use other subgroups, one must start with another decomposition of the space--time and write the wave function and so on.

The complete set of equations to be solved is the equation (\ref{e}), the WDW equation and the appropriate equation of matter field given by matter lagrangian. It must be noted here that solving the above mentioned equations is mathematically equivalent to solving the WDW equation and then using its decomposition to Hamilton--Jacobi equation and continuity equation, and extracting the Bohmian trajectories\cite{hor,hor1}. Writing the equations of motion, i.e. equation (\ref{e}) has formal importance and also we were able to investigate the symmetries of the theory using this approach. The quantum Einstein's equations are derived for a special metric (Robertson--Walker metric) previously\cite{vink}, but there neither any try is done to write the equations for a general metric, nor the symmetries are investigated.

At this end it is useful to present the conservation law. By getting the divergence of equation (\ref{e}), we have:
\begin{equation}
\nabla_\mu{\cal T}^{\mu\nu}=\frac{1}{\kappa}\nabla_\mu{\cal S}^{\mu\nu}
\label{f}
\end{equation}
Note that again we are restricted to the above mentioned subgroup.

\section{CONCLUSION}
We have shown in this paper that using the deBroglie--Bohm causal interpretation of quantum mechanics, one is able to obtain the appropriate Lie algebra of quantum gravity. It is shown that the presence of Bohm's quantum potential in the hamiltonian changes the classical algebra to clear spatial and temporal diffeomorphism algebra (this is valid weakly, that is only on the Bohmian trajectories.). We have also obtained Bohmian Einstein equations and shown that the quantum force modifies the dynamical parts while the non--dynamical parts are modified by the quantum potential itself. We also have shown explicitly that these quantum Einstein equations are form invariant under spatial and temporal diffeomorphisms. The important point is that these are done without any reference to some special mini super space or solution.

{\bf Acknowledgment}

The authors wish to thank Prof. T. Thiemann for useful discussions.

\appendix
{\bf Appendix}

In this appendix we shall calculate the hamiltonian constraint poisson bracket with itself. In evaluation of the integral in the equation (\ref{xax}), because of the relation (\ref{yay}) one needs an integral of the form:
\[\hfill \int d^3z F(q,p,N,N')\frac{\delta ^{(3)}{\cal R}}{\delta q_{ij}} \hfill (A.1) \]
So first we calculate the variation of the Ricci scalar with respect to the metric. Using the Palatini identity\cite{weinberg}, we have\footnote{Here for simplicity we drop $^{(3)}$ superscript and it is known that any quantity is a 3--quantity.}:
\[ \hfill \delta{\cal R}_{ij}=\frac{1}{2}q^{kl}\left ( \nabla_j\nabla_i\delta q_{kl} -\nabla_k\nabla_j\delta q_{li} -\nabla_k\nabla_i\delta q_{lj}+\nabla_k\nabla_l\delta q_{ij} \right ) \hfill (A.2) \]
so:
\[ \frac{\delta{\cal R}_{ij}(x)}{\delta q_{ab}(z)}=\frac{1}{2}\sqrt{q} q^{kl}\left ( \delta^a_k\delta^b_l \nabla_j\nabla_i \frac{\delta(x-z)}{\sqrt{q}}- \delta^a_l\delta^b_i \nabla_k\nabla_j \frac{\delta(x-z)}{\sqrt{q}} \right . \]
\[ \hfill \left . - \delta^a_l\delta^b_j \nabla_k\nabla_i \frac{\delta(x-z)}{\sqrt{q}} + \delta^a_i\delta^b_j \nabla_k\nabla_l \frac{\delta(x-z)}{\sqrt{q}} \right ) \hfill (A.3) \]
Therefore we have:
\[ \hfill \frac{\delta{\cal R}(x)}{\delta q_{ab}}= \frac{\delta (q^{ij}{\cal R}_{ij})(x)}{\delta q_{ab}}= -{\cal R}^{ab}\delta(x-z)+\sqrt{q}\left ( q^{ab}\nabla^2-\nabla^a\nabla^b\right )\frac{\delta(x-z)}{\sqrt{q}} \hfill (A.4) \]
Now using this identity in the relation (\ref{yay}), the only non vanishing terms in (\ref{xax}) are:
\[ \left \{ C(N), C(N')\right \} =\tilde{C}(N\vec{\nabla}N'-N'\vec{\nabla}N)-2\int d^3z d^3x \sqrt{q(x)q(z)}G_{ijkl}(z)p^{kl}(z) \times \hfill \] 
\[ \hfill \left ( N(x)N'(z)-N(z)N'(x)\right )\left ( q^{ij}(x)\nabla^2_x\frac{\delta(x-z)}{\sqrt{q}}- \nabla^i_x\nabla^j_x \frac{\delta(x-z)}{\sqrt{q}} \right )\hfill (A.5) \]
Integrating by part, we get:
\[ \left \{ C(N), C(N')\right \} =2\int d^3x \left ( \nabla_j(N\nabla_i N') 
- \nabla_j (N'\nabla_i N) \right ) p^{ij} \hfill \]
\[ +\int d^3x \sqrt{q}G_{ijkl}p^{kl}\left (N'\nabla^i\nabla^j N -N \nabla^i \nabla^j N'\right ) \hfill \] 
\[ \hfill -\int d^3x \sqrt{q}G_{ijkl}p^{kl}q^{ij}\left (N'\nabla^2 N-N\nabla^2N'\right )
 \hfill (A.6) \]
So that:
\[ \left \{ C(N), C(N')\right \} =2\int d^3x \left ( N\nabla_j\nabla_i N' -N' \nabla_j \nabla_i N \right ) p^{ij} \hfill \]
\[ +\int d^3x p^{kl}\left (N'\nabla_k\nabla_l N -N \nabla_k \nabla_l N' +N'\nabla_l\nabla_k N -N \nabla_l \nabla_k N' -q_{kl}\left ( N'\nabla^2 N -N\nabla^2 N'\right ) \right ) \hfill \] 
\[ \hfill +\int d^3x q_{kl}p^{kl}\left ( N'\nabla^2 N -N\nabla^2 N'\right )=0 \hfill (A.7) \]
So the desired result is obtained.

\end{document}